\documentclass[%
 reprint,
superscriptaddress,
 amsmath,amssymb,
 aps,
]{revtex4-2}

\usepackage{graphicx} 
\usepackage[utf8]{inputenc}
\usepackage{amsmath}
\usepackage{dcolumn}
\usepackage{bm}
\usepackage{siunitx}
\usepackage{cancel}
\usepackage[normalem]{ulem}
\usepackage{footnote}
\usepackage{float}
\usepackage{xcolor}
\usepackage{hyperref}

\hypersetup{
    colorlinks = true,
    linkcolor= blue,
    citecolor= blue,
    urlcolor= blue
}

\usepackage{cleveref}
\usepackage{natbib}
\usepackage{booktabs}
\usepackage{bbding}

\begin{document}

\title{Learning collective cell migratory dynamics\\ from a static snapshot with graph neural networks}

\author{Haiqian Yang}\thanks{hqyang@mit.edu}
\affiliation{Department of Mechanical Engineering, Massachusetts Institute of Technology, 77 Massachusetts Ave., Cambridge, MA 02139, USA}
\author{Florian Meyer}
\affiliation{Institute of Cell Biology and Immunology, University of Stuttgart, Allmandring 31, 70569 Stuttgart, Germany}
\author{Shaoxun Huang}
\affiliation{Department of Mechanical Engineering, Massachusetts Institute of Technology, 77 Massachusetts Ave., Cambridge, MA 02139, USA}
\author{Liu Yang}
\affiliation{Department of Computer Sciences, University of Wisconsin - Madison, Madison, WI 53706, USA}
\author{\\Cristiana Lungu}
\affiliation{Institute of Cell Biology and Immunology, University of Stuttgart, Allmandring 31, 70569 Stuttgart, Germany}
\author{Monilola A. Olayioye}
\affiliation{Institute of Cell Biology and Immunology, University of Stuttgart, Allmandring 31, 70569 Stuttgart, Germany}
\author{Markus J. Buehler}
\affiliation{Department of Mechanical Engineering, Massachusetts Institute of Technology, 77 Massachusetts Ave., Cambridge, MA 02139, USA}
\affiliation{Laboratory for Atomistic and Molecular Mechanics (LAMM), Massachusetts Institute of Technology, 77 Massachusetts Ave., Cambridge, MA 02139, USA}
\affiliation{Center for Computational Science and Engineering, Schwarzman College of Computing, Massachusetts Institute of Technology, 77 Massachusetts Ave., Cambridge, MA 02139, USA}

\author{Ming Guo}\thanks{guom@mit.edu}
\affiliation{Department of Mechanical Engineering, Massachusetts Institute of Technology, 77 Massachusetts Ave., Cambridge, MA 02139, USA}

\date{\today}

\begin{abstract}
Multicellular self-assembly into functional structures is a dynamic process that is critical in the development and diseases, including embryo development, organ formation, tumor invasion, and others. Being able to infer collective cell migratory dynamics from their static configuration is valuable for both understanding and predicting these complex processes. However, the identification of structural features that can indicate multicellular motion has been difficult, and existing metrics largely rely on physical instincts. Here we show that using a graph neural network (GNN), the motion of multicellular collectives can be inferred from a static snapshot of cell positions, in both experimental and synthetic datasets.

\vspace{35pt}
\end{abstract}

\maketitle

\section{Introduction}

Collective cell movement is crucial during many biological processes, such as embryogenesis, vascularization, cancer metastasis, and wound healing~\citep{keller2013imaging,xiong2014interplay,mcdole2018toto,kasza2019cellular,wang2020anisotropy,tang2022collective,han2020cell,kang2021novel,jeon2015human,zervantonakis2012three,xu2023geometry, skinner2021topological}. The inference of cell motion from multicellular structures not only can provide valuable information for understanding these complex processes, but also have a broad impact on medical and engineering fields such as histology, organ-on-chip, and 3D bio-printing for drug screening and disease modeling~\citep{kamm2018perspective}. Despite that modern optical microscopy has enabled visualizing the evolution of many living multicellular structures in real-time, the principles they follow to self-organize into a complex living structure is still a mystery~\citep{keller2013imaging,kamm2018perspective,trepat2018mesoscale,karsenti2008self}. 

\begin{figure*}[t]
    \centering
    \includegraphics[width=0.8\textwidth]{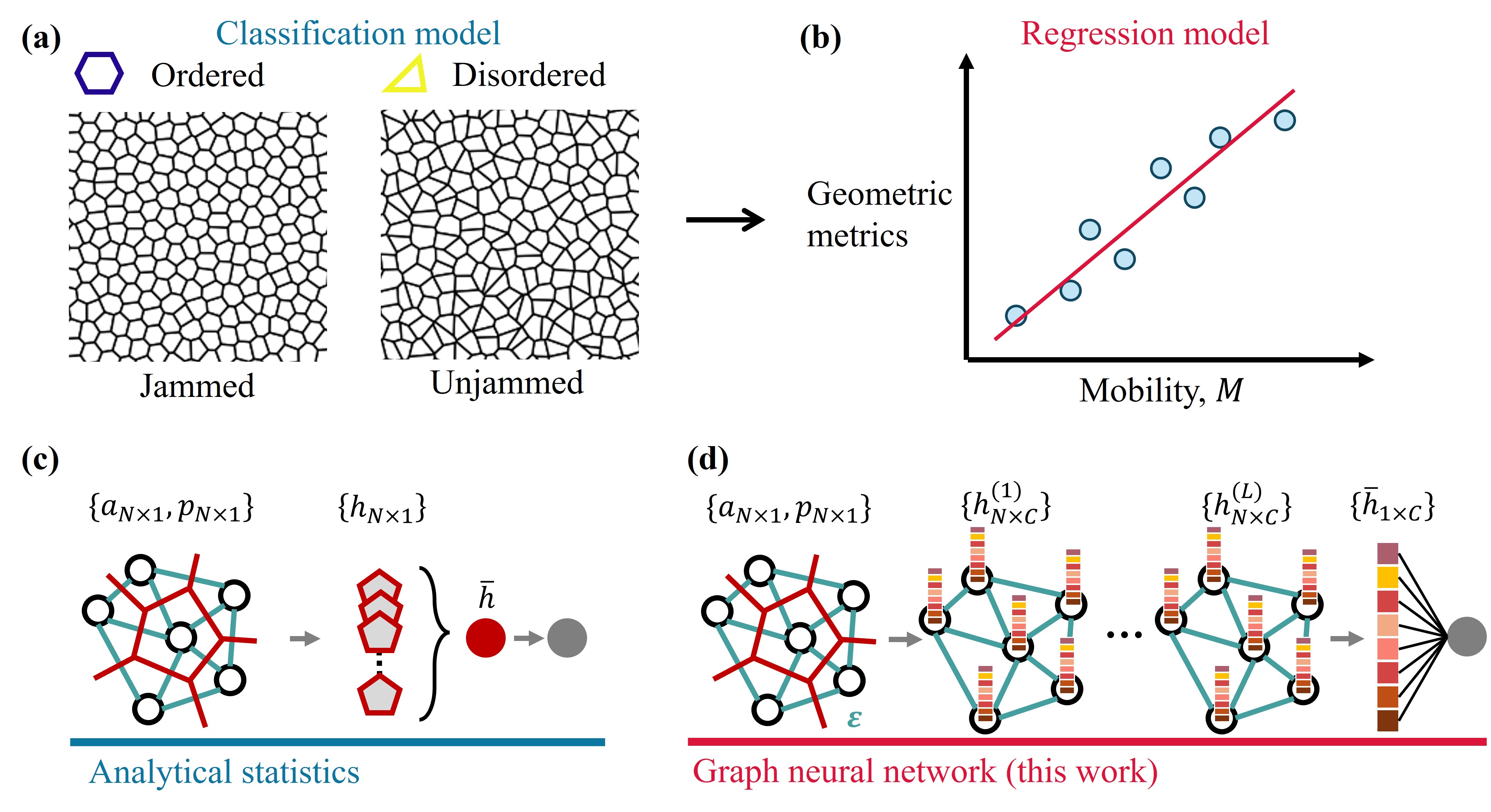}
    \caption{Infer cell motion from static multicellular configurations using graph neural network (GNN). (a) Classification model. Typically an ordered configuration corresponds to a jammed state while a disordered configuration corresponds to an unjammed state~\cite{bi2016motility,park2015unjamming}. Cell boundaries are shown here. (b) It remains elusive what geometric feature of the static snapshot can be used to build a regression model to make continuous predictions of collective cell migratory dynamics. (c) A schematic configuration is shown on the left, with the coordinates of the cells represented by circles, and cell boundaries marked in red. An analytical model typically calculates a geometrical feature for each cell $h_{N\times 1}$, from which an ensemble average $\bar{h}$ is taken to reflect the structural disorder that correlates with collective cell dynamics. (d) Here, this geometrical feature is replaced by a GNN with multiple graph convolutional layers and multiple hidden features $h_{N\times C}^{(l)}$ extracted from the training data through message-passing among neighboring cells via cell-cell adjacency $\mathcal{E}$, with $l \in \{1,2,...L\}$ the index of layers, and $C$ the number of channels. Cyan, the cell-cell adjacency $\mathcal{E}$. The ensemble average $\bar{h}_{1 \times C}^{(L)}$ is then taken on the multi-channel hidden features of the last graph convolutional layer, from which the prediction is calculated using a final linear layer.}
    \label{fig:shematics}
\end{figure*}

Over the past decade, collective cell movement has been primarily studied with non-equilibrium physics approaches~\cite{trepat2009physical,shaebani2020computational}. Particularly, recent theoretical and experimental studies have shown that cell collectives can undergo cell jamming and unjamming transitions, highlighting the relation between cell shapes and tissue fluidity~\cite{angelini2011glass,park2015unjamming,bi2015density,bi2016motility,atia2018geometric,wang2020anisotropy,lawson2021jamming,kang2021novel,yang2021configurational,huang2022shear}; a more ordered cell packing is more static, whereas a more disordered cell packing is associated with higher cell motility [Fig.~\ref{fig:shematics}(a)]. Nevertheless, these relations are categorical (i.e. jammed or unjammed), and it is still difficult to make continuous predictions of cell migratory dynamics [Fig.~\ref{fig:shematics}(b)]; for example, it was shown that cell shape is not sufficient to predict the average cell migration speed~\cite{kim2020unjamming}. {Given one static snapshot of two cell monolayers, at what resolution can we distinguish which cell monolayer is more dynamic than the other?} A continuous prediction of collective cell migratory dynamics can be potentially achieved by defining additional geometrical order parameters to refine the definition of structural disorder, but it is unclear how to systematically define a list of relevant parameters.

While classical mechanistic active-matter models rely on the assumptions of symmetries and constitutive relations, data-driven inference methods can potentially bypass some of these disadvantages~\cite{schmitt2024machine,cichos2020machine,bruckner2024learning,bruckner2019stochastic,romeo2021learning,bruckner2021learning,lachance2022learning,supekar2023learning,bhaskar2021topological,frishman2020learning,li2024irreversibility,gilpin2024generative}. Notably, recent developments in the graph-based deep neural networks (GNN)~\cite{corso2020PNA,kipf2016semi} provide an opportunity to develop data-driven models of complex systems, especially focused on models that take advantage of known discrete structural features. For instance, some studies have used graph-based modeling to capture complex multiscale material phenomena including dynamic properties, in diverse systems ranging from glass-forming liquids, proteins, spatial transcriptomics, crystalline materials, to spider webs~\citep{bapst2020unveiling,yang2022linking,guo2022rapid,vinas2023hypergraph,hu2024unsupervised,lu2023modeling}. Other work has focused on dynamic material phenomena using attention-based graph models in conjunction with denoising algorithms to model dynamic fracture~\cite{Buehler2022ModelingModel}. Besides, some recent studies have applied GNN to track microscopic motion including linking cell trajectories~\cite{pineda2023geometric}, and inferring rules that govern cell fate~\cite{yamamoto2022probing}. Nevertheless, to our knowledge, GNN models have not been applied to study the glassy dynamic behaviors of the mesoscale multicellular collectives; these multicellular systems are often represented with a vertex-model framework~\cite{bi2015density,bi2016motility,atia2018geometric,park2015unjamming,honda1978description} or network of cell-cell adjacencies~\cite{wang2020anisotropy,yang2021configurational,skinner2022topological,skinner2021topological}, making GNN models a suitable option to build data-driven algorithm to study their collective behaviors.

Here we propose to discover from static multicellular graph data the important spatial features that reflect collective cell migratory dynamics, using GNN models [Fig.~\ref{fig:shematics}(d)]. The models are trained and validated on both experimental and synthetic datasets, achieving accurate predictions. Furthermore, a series of ablation studies indicate cell geometries and their spatial interactions are important features that help the model with these predictions.

\begin{figure*}[t]
    \centering
    \includegraphics[width=0.95\textwidth]{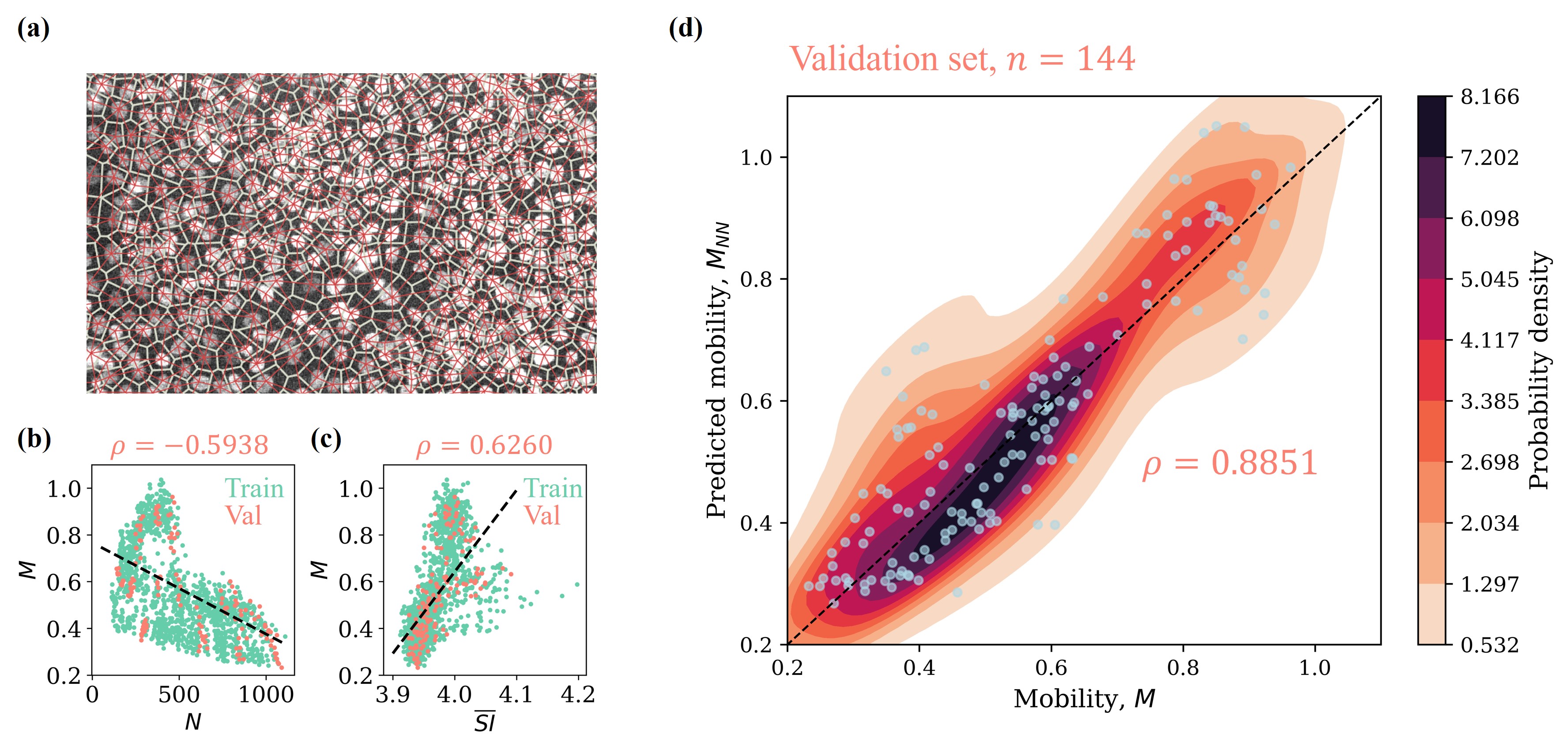}
    \caption{Inferring cell motion $\Tilde{M}(\mathbf{X},\mathcal{G})$ from a snapshot in the experimental dataset of MCF-10A cell monolayers cultured on 2D substrates. (a) A zoom-in view of a representative snapshot of the dataset. Nuclei are stained for cell tracking. Green, Voronoi edges. Red, Delaunay edges. (b) Mobility $M$ vs. cell number $N$ in the region of interest (ROI) with the dimension \SI{689.6}{\micro\meter} $\times$ \SI{492.6}{\micro\meter}, calculated in the train set (green) and the validation set (orange). The black dashed line is a fitted linear function of $M$ against $N$ in the validation set. (c) Mobility $M$ vs. median of cell shape index $\overline{SI}$ in the ROI, calculated in the train set (green) and the validation set (orange). The black dashed line is a fitted linear function of $M$ against $\overline{SI}$ in the validation set. (d) Mobility prediction $M_{NN}$ vs. ground truth $M$ in the validation set. The color map indicates probability density. The black dashed line is a reference line $M_{NN} = M$. In (b), (c), and (d), each dot indicates an individual snapshot (with $n=1,296$ individual snapshots in the train set and $n=144$ individual snapshots in the validation set). $\tau = 75$ minutes.}
    \label{fig:experiments}
\end{figure*}

\section{Problem setup}

The system is given by cell coordinates $\mathbf{r}_i(t) \in \mathbb{R}^{d}$, where $\mathbf{r}_i(t)$ is the Cartesian coordinates of the $i$th cell ($i=1, 2, ..., N$) at frame $t$, and $N$ is the total number of cells at frame $t$. We consider 2D systems ($d=2$) in this study. The dimensionless cell coordinates can be defined as $\mathbf{r}^*_i(t) = a_c^{-1/2}\mathbf{r}_i(t)$, where $a_c$ is a characteristic cell area (See Materials and Methods for details of $a_c$).

\textbf{Mobility output.}---
We quantify the migratory dynamics by the dimensionless average traveled distance over a lag time $\tau$, which we define as the mobility, 
\begin{equation}
    M(\tau) = \langle \triangle {r^*}^2(\tau) \rangle^{\frac{1}{2}},
\end{equation}
where $\langle \triangle {r^*}^2(\tau) \rangle$ is the dimensionless ensemble-averaged radial mean squared displacement. $M$ is a measure of the average distance traveled relative to a characteristic cell size.

\textbf{Graph input.}---
At time point $t=t_s$, from one snapshot of the dimensionless cell centroid coordinates $\mathbf{r}^*_i(t_s)$, Delaunay triangulation can be used to construct the cell-cell adjacency. Each input graph $\mathcal{G}$ is given by [Fig.~\ref{fig:shematics}(d)]
\begin{equation}
    \mathcal{G} = (\mathcal{N}, \mathcal{E})\, ,
\end{equation}
which contains cell centroids (nodes) $n_i \in \mathcal{N},\, \mathcal{N}=\{1,2, \ldots, N\}$, and cell-cell adjacencies (undirectional Delaunay edges) $\mathbf{e}_k = (n_i,n_j) \in \mathcal{E},\, \mathcal{E} \subseteq \mathbb{Z}^{2 \times E}$, with $E$ the number of edges.

Each node is described by a feature vector $\mathbf{c}_i \in \mathbb{R}^{1\times C}$, with $C$ the dimension of input nodal features. Because the output $M$ is invariant, we construct this input feature vector from invariant geometric quantities of the Voronoi tessellation. Two invariants, area and perimeter are shown to be particularly important in regulating tissue fluidity~\cite{bi2016motility,park2015unjamming}. Therefore, while higher order moments can be considered, here we consider $C=2$ and
\begin{equation}
    \mathbf{c}_i = (a_i,p_i)
\end{equation}
with $a_i$ the dimensionless area and $p_i$ the dimensionless perimeter of the Voronoi cell corresponding to the node $n_i$. These feature vectors are organized in the feature matrix $\mathbf{X} = [\mathbf{c}_1\,\mathbf{c}_2\, ... \mathbf{c}_N] \in \mathbb{R}^{N \times C}$.

\textbf{Task.}---
The goal is to identify a neural network approximation of the following function
\begin{equation}\label{Eq:Mtau}
    M(\tau) = \Tilde{M}(\mathbf{X},\mathcal{G}),
\end{equation}
which relates a static snapshot of the system $(\mathbf{X}, \mathcal{G})$ at time point $t=t_s$ with the mobility $M(\tau)$.

\section{Results}
\subsection{Performance on experimental dataset}

To test this approach in experimental multicellular systems, we first perform experiments to perturb cell motility and their static configurations. We use the non-tumorigenic human breast epithelial cell line MCF-10A as our model system. Previous studies have shown that epidermal growth factor (EGF) withdrawal can induce motility arrested phase~\cite{leggett2019motility}, while transforming growth factor $\beta$ (TGF$\beta$) treatment can enhance cell motility through an Epithelial-Mesenchymal-Transition (EMT)-like process~\cite{seton2004cooperation,hosseini2020emt}. Besides the biochemical cues, biophysical environment such as cell density not only results in glassy behaviors~\cite{yang2021configurational,angelini2011glass}, but also affects the formation of cell-cell junctions and contact inhibition mechanisms that maintain tissue homeostasis~\cite{brenner2024repeat,hanahan2011hallmarks}. Therefore, to generate a comprehensive dataset that covers a wide range of static structures and cell motility, we perform time-lapsed imaging of 2D MCF-10A monolayers under different conditions, including EGF withdrawal, TGF$\beta$ treatment, and the control group with no treatment, each at a series of cell seeding densities (See {\color{blue}Supplemental Material} for details of the experiments). We create an experimental dataset of 240 independent 16-hour time-lapsed videos of MCF-10A cell monolayers. Among the 240 videos, 216 videos are used to construct the training set, and 24 videos are used to construct the validation set. We select 6 well-separated frames of each video, each as one entry in the datasets, resulting in a total of $1,296$ data entries for training, and $144$ data entries for validation (see Materials and Methods for details regarding preparing the graph from raw videos).

\begin{figure*}[t]
    \centering
    \includegraphics[width=0.8\textwidth]{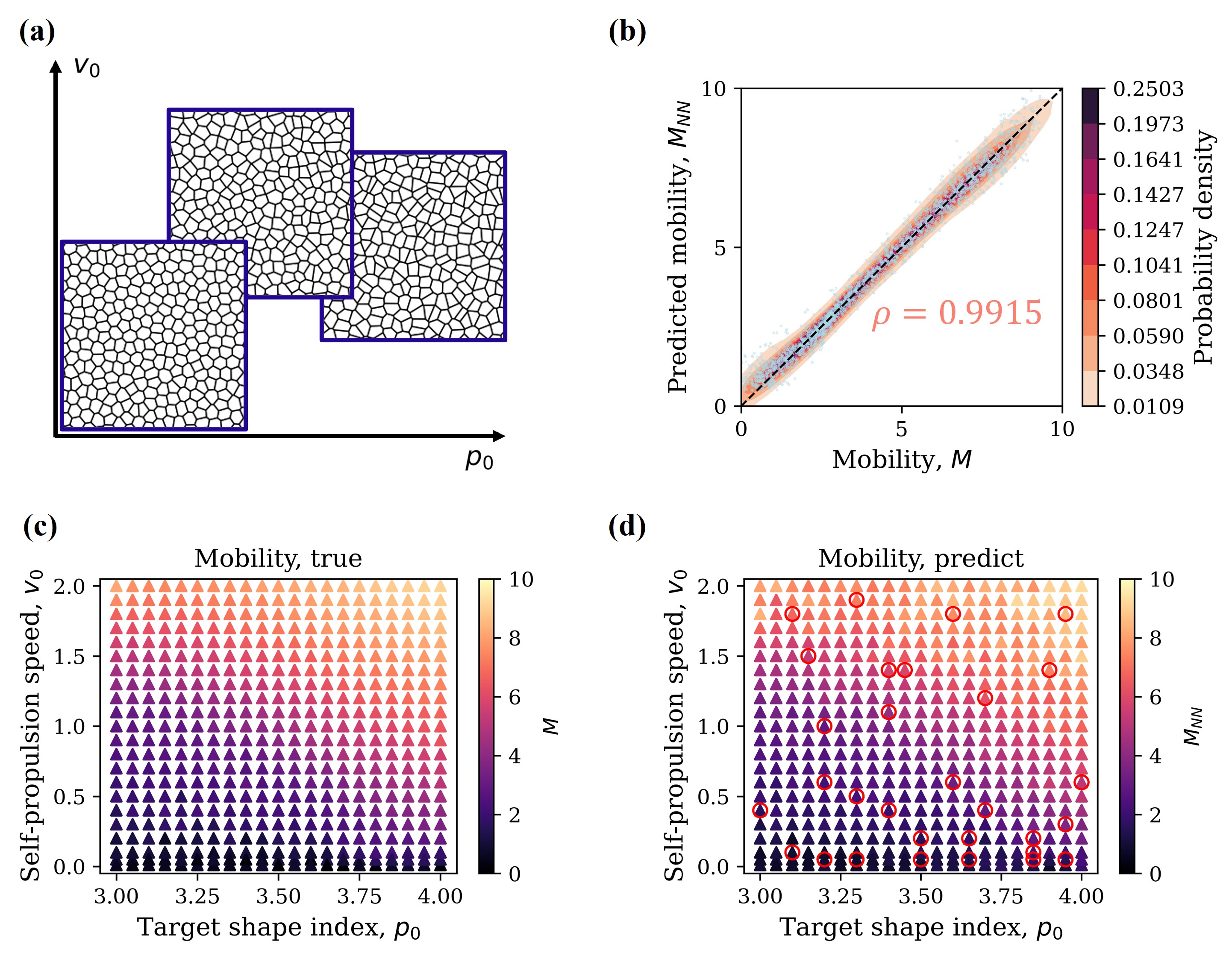}
    \caption{The GNN model approximates $\Tilde{M}(\mathbf{X},\mathcal{G})$ in simulated cell monolayers. (a) Representative input configurations. The cell boundaries are shown. (b) Prediction vs. ground truth. The color map indicates probability density, and each dot indicates an individual snapshot. The black dashed line is a reference line $M_{NN} = M$. (c) Ground-truth mobility landscape. The color map indicates the ground-truth mobility. (d) Predicted mobility landscape. Red markers, the state points ($N_s=30$) provided for training. The color map indicates the predicted mobility. The dimensionless time $\tau=10$. }
    \label{fig:simulation}
\end{figure*}

The mobility $M$ has a wide distribution ({\color{blue}Supplemental Material} Fig.~S2); on both train and validation sets, the apparent geometrical features such as cell number $N$ in the field of view or median cell shape index $\overline{SI}$ are only weakly correlated with the mobility $M$ [Fig.~\ref{fig:experiments}(b) and (c)], with Pearson correlation of $-0.5938$ and $0.6260$ respectively on the validation set ($-0.5728$ and $0.5599$ on the whole dataset), suggesting the need of additional structural features to further distinguish these systems.

Interestingly, the trained GNN model achieves highly accurate prediction of cell mobility $M_{NN}$ from a static snapshot of the system, with a Pearson correlation $\rho = 0.8851$ between the prediction and ground truth [Fig.~\ref{fig:experiments}(d)]. Furthermore, we train the model on cell mobility at different lag times $\tau$, and we find the trained models generate consistently accurate predictions across a range of $\tau$ from 30 minutes to 2 hours ({\color{blue}Supplemental Material} Fig.~S4 and Fig.~S5).

\subsection{Performance on synthetic dataset}\label{Learn}

To further test the GNN model, we use a synthetic dataset with continuously varying cell-cell interaction and self-propulsion speed at a constant cell number density~\cite{bi2016motility,yang2021configurational}. Each simulation in the dataset is performed with different target shape index $p_0$ and self-propulsion speed $v_0$, and the dataset consists of 462 distinct sets of configurations with different $(p_0,v_0)$ [Fig.~\ref{fig:simulation}(c)]. In this case, we seek to use a GNN model to interpolate the mobility from the graphs, without information about $p_0$ and $v_0$. To do so, we randomly select a small number of state points to train GNN models and validate the trained models on the whole dataset. Here, from 30 randomly selected state points, the GNN model achieves high accuracy with a Pearson correlation $\rho = 0.9915$ and recovers the mobility landscape [Figs.~\ref{fig:simulation}(c\&d)]. Furthermore, we train and validate models at various time scales $\tau$, and we find that the models generate accurate predictions consistently ({\color{blue} Supplemental Material} Fig.~S6 and Fig.~S7). Changing the number of state points randomly selected for training, we observe that, with 5 state points provided for training, the GNN model can already achieve a correlation of $\rho=0.9159\pm 0.0445$, while 10 state points can improve it to $\rho = 0.9552\pm 0.0264$, 20 state points improve it to $\rho=0.9760\pm 0.0146$, and 40 state points can increase it to $0.9900 \pm 0.0034$ ({\color{blue}Supplemental Material} Table.~S2). Providing state points on regular mesh grids in general generates slightly better performance compared to random selection ({\color{blue}Supplemental Material} Table.~S4 and Fig.~S8).

\bigskip

To summarize, the predictions achieved by the GNN models indicate that static multicellular configurations contain critical features that can be utilized to predict multicellular dynamics. While it has been challenging for classical mechanistic models to perform regression tasks between static graphs and tissue dynamics, graph-based deep neural networks provide an alternative solution.

\subsection{Ablation study}

\begin{table}[h!]
    \centering
    \begin{tabular}{p{1.5cm}|p{1.5cm}|p{1.5cm}|p{1.5cm}|p{1.5cm}}
        \toprule
        \multicolumn{2}{c}{Features}& \multicolumn{3}{c}{Type of information}  \\
        \midrule
        Node        &  Edge         & Geometric          & Neighbor      & Invariant        \\
        \midrule
         $(a,p)$    &  $\mathbf{e}$ &    \Checkmark      & \Checkmark    &  \Checkmark       \\
         $(a,p)$    &  -            &    \Checkmark      &               &  \Checkmark       \\
         -          &  $\mathbf{e}$ &                    & \Checkmark    &  \Checkmark       \\
         $(x,y)$    &  $\mathbf{e}$ &    \Checkmark      & \Checkmark    &                   \\
         \bottomrule
    \end{tabular}
    \caption{Four typical inputs are compared, with different combinations of information provided.}
    \label{tab:table1}
\end{table}

\begin{figure}[h!]
    \centering
    \includegraphics[width=0.4\textwidth]{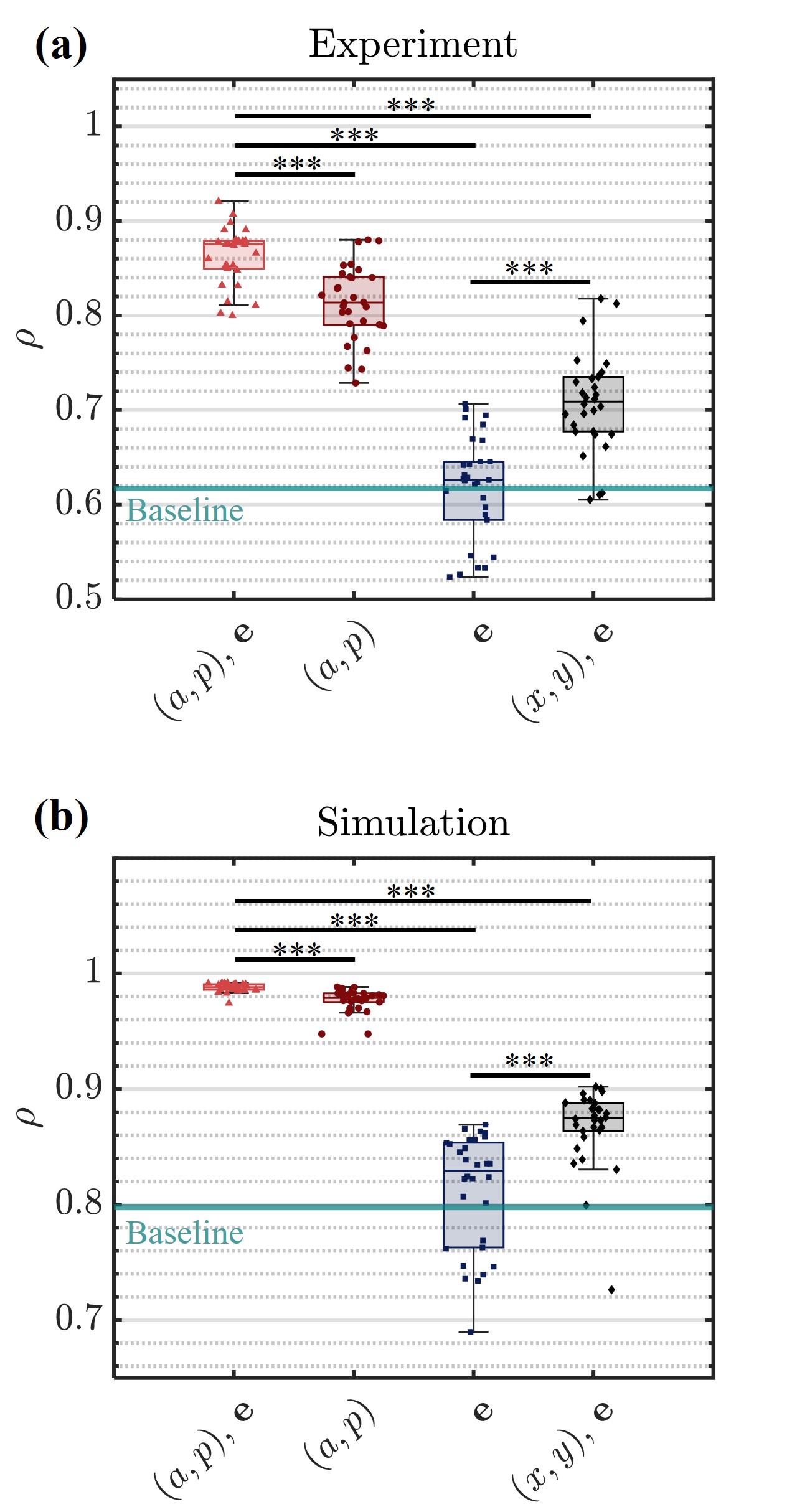}
    \caption{Comparing the performance with different inputs. (a) On the experiment dataset, 10-fold cross-validation is performed, and each fold is repeated with 3 different model initializations at different random seed numbers. Each point represents the performance of an independent model (30 models). (b) On the synthetic dataset, cross-validation is performed by randomly selecting the training set ($N_s=30$ state points) and model initialization. Each point represents the performance of an independent model (30 models). The statistical significance level is determined using One-way ANOVA tests and post-hoc pairwise Welch's $t$-test with Bonferroni's correction for multiple comparisons. ***: p-value $<0.001$. Implementation details of ablation and cross-validation are provided in the Materials and Methods.}
    \label{fig:compare}
\end{figure}

Intuitively, the GNN model achieves good performance because it can capture nonlinear spatial interaction. To understand what information is indeed important for its performance, we further perform cross-validation with partial input information ablated or altered in both the training set and the validation set (Table~\ref{tab:table1}, Fig.~\ref{fig:compare}, see Materials and Methods for implementation details).

By default, we provide area and perimeter $(a,p)$ as node features together with cell-cell adjacency $\mathbf{e}$; this achieves a Pearson correlation of $\rho = 0.8637 \pm 0.0299$ in the experimental dataset [Fig.~\ref{fig:compare}(a)], significantly better than the baseline value $\rho = 0.6173$ (determined using linear regression of $M$ against cell number $N$ and median cell shape index $\overline{SI}$ over the full dataset, details in the Materials and Methods). Removing edges reduces $\rho$ to $0.8134 \pm 0.0399$, suggesting that neighborhood features are useful in predicting cell dynamics. This is consistent with the known flocking behaviors of cells, which have been modeled by the Vicsek model and others~\cite{vicsek1995novel,trepat2018mesoscale}; recent studies also show that cell neighborhood information such as cell-cell tension affects collective cell migratory dynamics~\cite{herrera2023tissue,perez2023tension}. Message-passing in the neighborhood as defined by the edges across multiple graph convolutional layers allows the model to consider structural order at different ranges, which has been observed in cell collectives~\cite{tang2024cell,cislo2023active,puliafito2012collective}. The experimental task appears to be more difficult than the synthetic dataset; on the latter, even without the edge information, the model already reaches a correlation of $0.9766 \pm 0.0098$. Providing the additional edge information only has a marginal gain, increasing the correlation to $\rho = 0.9876 \pm 0.0037$ [Fig.~\ref{fig:compare}(b)]. In the vertex-model framework, the cell-cell tension linearly depends on the perimeter of two neighboring cells~\cite{huang2022shear,nguyen2024origin}, which constitutes a simple neighborhood feature.

Furthermore, fully removing the node feature decreases $\rho$ to $0.6124 \pm 0.0675$ in the experimental dataset, and $0.8121 \pm 0.0504$ in the synthetic dataset. This is consistent with mechanistic theories showing cell shape is closely related to tissue fluidity~\cite{bi2016motility}. Interestingly, adding cell positions $(x,y)$ back instead as the node feature can only partially rescue the performance, with $\rho$ to $0.7072 \pm 0.0519$ in the experimental dataset, and $0.8668 \pm 0.0350$ in the synthetic dataset. While the raw cell coordinates may provide more information, the model benefits from pruning redundant input using symmetry arguments. In this case, the output $M$ is an invariant quantity, and providing invariants as input enforces the rotational symmetry.

This ablation study suggests that cell geometries and their spatial interactions contain important features that can be used for accurate inference of tissue-level cell mobility. While classical mechanistic models heavily rely on intuitions and assumptions to distill a handful of structural metrics, GNN models provide an alternative solution to extract structural features from data that are relevant to tissue dynamics.

\section{Discussions}

Here we apply a graph neural network (GNN) to approximate the relation between multicellular static structures and collective cell dynamics. Our results suggest that the static structures of cell monolayers contain sufficient information to infer cell migratory dynamics with reasonable accuracy; using ablation studies, we demonstrate the relative importance of different input features for the model performance. We present that GNN is a promising tool that can be used to approximate relations between multicellular graph data and collective cell migratory dynamics, which are otherwise difficult to express using analytical formulas with predetermined structural order parameters.

Besides cellular geometries, recent studies have shown that collective cell dynamics are also regulated by cellular biochemical identities such as actomyosin activities~\cite{herrera2023tissue}, tension remodeling~\cite{perez2023tension}, and the extracellular-signal-regulated kinase (ERK) pathway~\cite{hino2020erk}, and other studies have shown the role of nematic order~\cite{armengol2023epithelia}. GNN models could allow studying their interactions together with cell geometries by further concatenating these cellular identities as additional dimensions of the input features. Moreover, in several recent studies, inference of dynamic equations has been achieved on single-cell trajectories~\cite{bruckner2019stochastic,bruckner2020inferring,bruckner2021learning,romeo2021learning}. It will be interesting to combine GNN with these methods to infer the forces within dense active matter and living tissues such as epithelia.

We envision that geometric deep-learning algorithms will provide a suitable tool for studying diverse physical phenomena in mesoscale multicellular living systems and dense active matter. We further discuss some of the possibilities in the {\color{blue}Supplemental Material}.

\section{Materials and Methods}\label{sec:Methods}

\textbf{Experimental dataset.}---
Experimental and data pre-processing details can be found in the {\color{blue}Supplemental Material}. Briefly, we use the non-tumorigenic human breast epithelial cell line MCF-10A to create the experimental dataset. The cells are cultured in a glass bottom 96-well plate (Greiner, No.~655891) coated with collagen R (Serva, No. 47254.01), and the nuclei are stained (SPY650-DNA) for cell tracking. The cells are imaged using WiScan Hermes High Content Imaging System (Idea Biomedical) every 3 minutes for 16 hours. To perturb both cell mobility and their static configuration, three types of treatment conditions (EGF withdrawal, TGF$\beta$ treatment, and control) and 4 initial cell seeding number densities are used, and 20 independent locations of each condition are imaged, resulting in a total of 240 independent time-lapsed videos, including both confluent and non-confluent cell monolayers, covering a broad range of monolayer structures and cell mobility. Among the 240 videos, 216 videos are used to create the train set, and 24 videos are used to create the validation set.

Cell tracking is performed with TrackMate~\cite{ershov2022trackmate} and a customized MATLAB script adapted from the particle tracking algorithm~\cite{crocker1996methods}. Multiple frames at 1-hour intervals are selected from each video, resulting in a train set of 1,296 graphs, and a validation set of 144 graphs. The field of view (FOV) of the raw videos has a dimension of \SI{802}{\micro\meter} $\times$ \SI{599}{\micro\meter}. The characteristic length scale $a_c^{1/2}$ is calculated as the $a_c^{1/2} = \sqrt{\frac{A_{FOV}}{\langle N \rangle}} = \SI{24.6}{\micro\meter}$ for our dataset, with $A_{FOV}$ the area of the FOV, and $\langle N\rangle$ the average cell number across the full dataset consisting of $1440$ graphs. For the input graphs $\mathcal{G}$, we filter out the distorted tiles at the boundary of the FOV, and the final region of interest (ROI) has a dimension of \SI{689.6}{\micro\meter} $\times$ \SI{492.6}{\micro\meter}.

\textbf{Synthetic dataset.}---
The simulation dataset was created with Self-Propelled Voronoi simulations as previously described in~\cite{bi2016motility,yang2021configurational}. The dataset contains steady-state cell coordinates at 462 distinct state points at different $(p_0, v_0)$ as shown in Fig.~\ref{fig:simulation}(c). Unless otherwise noted, we provide 7 distinct snapshots per state point to construct the full dataset ($n=3,234$), and we provide 30 randomly selected state points as marked in Fig.~\ref{fig:simulation}(d) for training ($n=210$). Model performance with a different number of snapshots per state point or different state points provided for training is provided in the {\color{blue}Supplemental Material}.

\textbf{Training GNN.}---Details of the GNN model architecture are provided in the {\color{blue}Supplemental Material}. The Principal Neighbourhood Aggregation (PNA) graph convolutional layer is employed~\cite{corso2020PNA}, and the models are implemented with PyTorch Geometric~\cite{fey2019graph}. The mean squared error loss function is used throughout the paper.

\textbf{Pearson correlation.}---
Throughout this study, the Pearson correlation factor $\rho$ is used to evaluate and compare the performance of the models. The Pearson correlation factor $\rho$ between the ground truth and prediction is calculated as
\begin{equation}
    \rho = \frac{\sum (M_{NN}-\langle M_{NN}\rangle)(M-\langle M \rangle)}
    {\sqrt{\sum (M_{NN}-\langle M_{NN} \rangle)^2\sum (M-\langle M \rangle)^2}}\,,
\end{equation}
where the summation is taken over the whole validation set. $\langle {M}_{NN} \rangle$ and $\langle M \rangle$ denote average neural-network prediction and average ground truth respectively.

\textbf{Ablation study.}---
To ablate the node feature, we replace all elements in the node feature matrix $\mathbf{X}$ by 0. To ablate the edges, we replace the edge list $\mathcal{E}$ with only self-loops $(n_i,n_i)$, where $i=1, 2, ..., N$.

\textbf{Cross-validation on the experimental dataset.}---
Since 20 independent locations are images at all conditions, we split the whole dataset by their location index into 10 subgroups, each containing 2 independent locations of each of the 3 treatment conditions and 4 cell seeding densities, with 72 graphs per location and 144 graphs per subgroup. We perform a 10-fold cross-validation. In each fold, we select 1 subgroup as the validation set and use the other 9 subgroups as the training set; each fold is repeated with model initialization using 3 different random seed numbers.

\textbf{Cross-validation on the synthetic dataset.}---
With each random seed number, among the 462 state points, a total of 30 sets of training data are selected. 30 state points are randomly selected for each set of training data, and the ablation study is performed on each training set.

\textbf{Baseline model.}---
Previous theoretical and experimental studies have identified the median of cell shape index $\overline{SI}$ as a shape order parameter that defines the jamming-unjamming phase boundary of epithelia~\cite{bi2015density,park2015unjamming}. In the experimental dataset, the baseline performance is determined using a two-variable linear regression model of the cell mobility $M$ as a function of the median cell shape index $\overline{SI}$ and the cell number $N$ in the ROI, over the full dataset consisting of $1440$ graphs. The synthetic dataset is at a constant cell number density, and the baseline performance is determined using a linear regression model of the cell mobility $M$ as a function of the median of the cell shape index $\overline{SI}$, over the full dataset consisting of 462 state points.

\textbf{Statistics.}---
Performance is reported as mean $\pm$ standard deviation. Statistic significance is determined using one-way ANOVA (Analysis of Variance) tests with post-hoc pairwise Welch's $t$-test with Bonferroni's correction for multiple comparisons.

\section*{Data availability}

Our codes and trained model weights are available at

\url{https://github.com/GuoLab-CellMechanics/GNN-collective-cell-dynamics}

Our experimental dataset is available at

\url{https://doi.org/10.5281/zenodo.13988939}

\section*{Author contribution}
H.Y., M.B., and M.G. conceptualized this study. H.Y. performed this study. H.Y., L.Y., and M.B. wrote the machine learning algorithm. F.M., C.L., and M.A.O. collected the experimental cell monolayer data, and H.Y. and S.H. performed cell tracking. H.Y., M.B., and M.G. wrote the paper.

\section*{Acknowledgements}
We thank Dapeng Bi for providing the synthetic dataset. We thank Roger Kamm, Zhenze Yang, Anh Nguyen, Brendan Unikewicz, and Audrey Parker for the helpful discussions. The authors gratefully acknowledge the Technology Platform “Cellular Analytics” of the Stuttgart Research Center Systems Biology. MG acknowledges support from NIH (1R01GM140108). MJB acknowledges support from NIH (5R01AR077793). CL and MAO acknowledge supports from the Baden-Wuerttemberg Ministry of Science, Research and Arts. CL, MG and MAO received support from a Massachusetts Institute of Technology International Science and Technology Initiatives–Germany seed fund.

\bibliography{bibliography}

\end{document}


\title{Supplemental Material for\\ Learning collective cell migratory dynamics\\ from a static snapshot with graph neural networks}

\author{Haiqian Yang}\thanks{hqyang@mit.edu}
\affiliation{Department of Mechanical Engineering, Massachusetts Institute of Technology, 77 Massachusetts Ave., Cambridge, MA 02139, USA}
\author{Florian Meyer}
\affiliation{Institute of Cell Biology and Immunology, University of Stuttgart, Allmandring 31, 70569 Stuttgart, Germany}
\author{Shaoxun Huang}
\affiliation{Department of Mechanical Engineering, Massachusetts Institute of Technology, 77 Massachusetts Ave., Cambridge, MA 02139, USA}
\author{Liu Yang}
\affiliation{Department of Computer Sciences, University of Wisconsin - Madison, Madison, WI 53706, USA}
\author{\\Cristiana Lungu}
\affiliation{Institute of Cell Biology and Immunology, University of Stuttgart, Allmandring 31, 70569 Stuttgart, Germany}
\author{Monilola A. Olayioye}
\affiliation{Institute of Cell Biology and Immunology, University of Stuttgart, Allmandring 31, 70569 Stuttgart, Germany}
\author{Markus J. Buehler}
\affiliation{Department of Mechanical Engineering, Massachusetts Institute of Technology, 77 Massachusetts Ave., Cambridge, MA 02139, USA}
\affiliation{Laboratory for Atomistic and Molecular Mechanics (LAMM), Massachusetts Institute of Technology, 77 Massachusetts Ave., Cambridge, MA 02139, USA}
\affiliation{Center for Computational Science and Engineering, Schwarzman College of Computing, Massachusetts Institute of Technology, 77 Massachusetts Ave., Cambridge, MA 02139, USA}

\author{Ming Guo}\thanks{guom@mit.edu}
\affiliation{Department of Mechanical Engineering, Massachusetts Institute of Technology, 77 Massachusetts Ave., Cambridge, MA 02139, USA}

\date{\today}

\maketitle

\appendix
\onecolumngrid

\renewcommand\thefigure{S\arabic{figure}}   
\renewcommand{\thetable}{S\arabic{table}}
\setcounter{figure}{0}
\setcounter{table}{0}

\tableofcontents
\clearpage

\subsection{Details of the graph neural network model}

\begin{figure*}[h]
    \centering
    \includegraphics[width=0.95\textwidth]{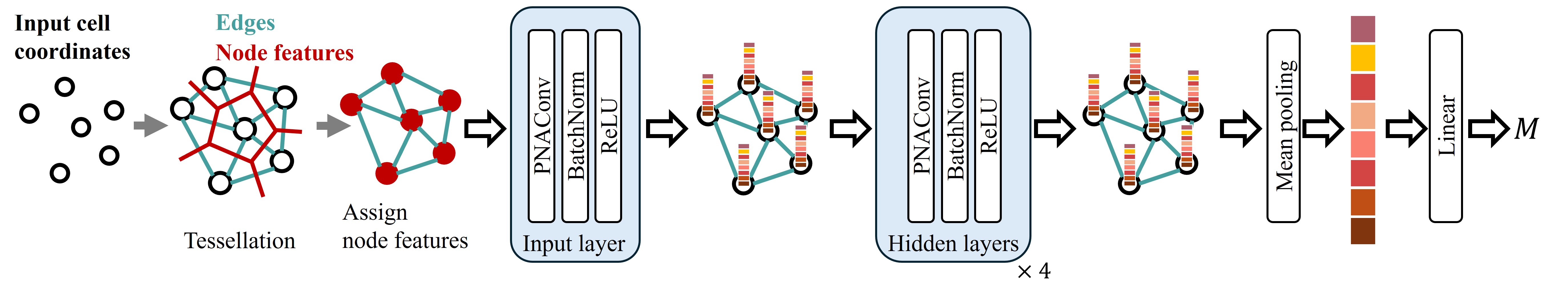}
    \caption{Model architecture.}
    \label{fig:modelArchitecture}
\end{figure*}

From cell coordinates, Voronoi tessellation is performed using \texttt{scipy.spatial.voronoi}, and the area $a$ and perimeter $p$ of the Voronoi cells are assigned as nodal features $\mathbf{X}$; Delaunay tessellation is performed using \texttt{scipy.spatial.delaunay} to get the cell-cell adjacency as the graph edges $\mathcal{E}$. Unless otherwise specified, the network consists of 5 PNA convolutional layers~\cite{corso2020PNA} with 36 channels and 3 towers (model performance using different choices of hyperparameters is shown in Table.~\ref{tab:HyperParam}). ReLU activation function is used and batch norm layers are applied after each PNA layer. Global mean pooling and another linear layer are applied after the last PNA layer to transform to the output dimension (Fig.~\ref{fig:modelArchitecture}). In the PNA convolutional layers~\cite{corso2020PNA}, we use  \texttt{[sum, mean, std, max, min]} aggregators, along with \texttt{[identity, amplification, attenuation]} scalers. AdamW~\cite{loshchilov2017decoupled} with weight decay 1e-3 is used. We use learning rate 1e-5 for the experimental dataset and 1e-4 for the synthetic dataset. All models are trained for 500 epochs. All models are trained and validated with a NVIDIA T4 or L4 GPU.

\clearpage
\subsection{Additional discussions}

\subsubsection{Inference and generative tasks}

While continuum hydrodynamic differential equations for physics systems, including active matter, are classically developed through symmetry argument and approximate free energy function with polynomial expansion, a significant effort has been made over the past decade to use neural networks for inferring the governing rules and equations from data~\cite{supekar2023learning,rudy2017data,brunton2016discovering}, and to use physics-informed neural networks (PINN) to accelerate computation~\cite{karniadakis2021physics}. Similar opportunities exist in mesoscale dense active matter research, where the free energy function is usually assumed to be functions of certain invariant geometric quantities of cells (e.g. area and perimeter) in quadratic forms for simplicity. Designing a data-driven method to infer the forms and parameters of these terms in the governing equations from experimental data has remained a challenge. Doing so will be useful to refine the governing equations and free energy functions that have been widely used to simulate these mesoscale multicellular systems. For this task, compared to CNN which is more suitable for continuous fields, GNN is ideal for mesoscale active matter that has been widely described with Voronoi and Delaunay graphs for a long time~\cite{trepat2018mesoscale,bi2016motility,honda1978description}.

Bridging discrete multicellular graphs and continuous fields, recent studies have also revealed an important correlation between multicellular structures and tissue stresses in epithelia in both experiments~\cite{saraswathibhatla2022coordination} and simulations~\cite{yang2017correlating}. Generating stress fields from graph data has been achieved using GNN on crystalline structures on synthetic datasets~\cite{yang2022linking}. It is interesting to apply similar frameworks for inference of collective cell traction and tension fields from multicellular graphs. While invariance is enforced in the current study because the mobility measure $M$ is an invariant quantity, in learning vector and tensor fields such as stresses, equivariant GNN structures~\cite{musaelian2023learning} need to be considered.

\subsubsection{Multi-body interactions in dense active matter}

To further understand how graph-based learning might have gained additional insights into multicellular dynamics, we can consider a multi-body system whose dynamic equation is $\frac{d\xv_i}{dt} = \mu \mathbf{F(\{\xv_i\})} + v_0 \boldsymbol{\zeta}$, where $\xv_i$ is the Cartesian coordinates, $t$ is time, $\Fv$ is the interaction force that could in principle depend on all the degrees of freedom $\xv_i$, $\boldsymbol{\zeta}$ is a unit noise term, $\mu$ is a coefficient and $v_0$ is the self-propulsion speed, $i=1,2,...N$ is the index of the $i$th cell. While this is still an oversimplification of multicellular living systems, we can gain some insight into how learning might benefit from a graph-based construction.

Here the multi-body force $\Fv(\{\xv_i\})$ is the unknown, and we are given a number of observations of $\{\xv_i\}$. The task of either mechanistic models or deep learning models can be regarded as inferring either an analytical approximation or a neural network approximation for $\Fv(\{\xv_i\})$, which is highly complex in living tissues. A straightforward choice is to expand $\Fv(\{\xv_i\})$ as a sum of interaction terms~\cite{bruckner2024learning}:
$\Fv(\{\xv_i\}) \simeq  \sum_i \Fv^{(1)}(\xv_i) 
+ \sum_i \sum_j \Fv^{(2)}(\xv_i,\xv_j) 
+ \sum_i \sum_j \sum_k \Fv^{(3)}(\xv_i,\xv_j, \xv_k)+...$ 
In recent studies, several inference models have been constructed for active matter under this decomposition and a focus has been on estimating the two-body term $\sum_i \sum_j \Fv^{(2)}(\xv_i,\xv_j)$. The underlining assumption is that $|\Fv^{(n>2)}| \ll |\Fv^{(2)}|$. This assumption is generally true in classical systems, but in dense active matter such as living tissues, such interactions can be multi-body in nature, meaning that $|\Fv^{(n>2)}|$ is not necessarily negligible compared to $|\Fv^{(2)}|$, but rather depends on the relative cell-cell interactions across many cells. Graph-based learning provides an alternative option through estimating $\Fv$ instead from the neighborhood of the cells represented as ``graphs". Notably, in more realistic situations where $\Fv$ is not only a function of cell positions $\{\xv_i\}$, but also a function of other multi-omics $\{\mathbf{o}_i\}$, GNN models are flexible enough to concatenate the information together as input embedding.

\subsubsection{Efficient data representation, Multicellular Data Bank (MDB), and Large Multicellular Graph Model (LMGM)}

Furthermore, we propose that a collaborative effort can be made to create a multicellular data bank (MDB) from which it will be possible to construct a large multicellular graph model (LMGM) for general-purposed predictions of the dynamic behaviors of multicellular living systems.

This century witnessed the fast growth of multicellular data for a variety of tissues and organs, yet it has been difficult to identify a universal model (either theoretical or computational) that can be predictive of the organization and dynamics of multicellular systems. It is still unclear what a standard representation of multicellular data is. Learning from data, the excellent performance of graph-based deep neural networks proves that multicellular graphs contain important hidden information that determines multicellular dynamics. On the other hand, from a technical aspect, the raw multicellular data are high-dimensional, typically z-stacked (3D), time-lapse (dynamical), and multi-channel (multi-omics). The data size for a single biological sample at acceptable resolution can typically be on the order of 1-10 gigabytes. A large-scale deep neural network based on video representation greatly exceeds the current data process, transfer, and storage capacity. For the purpose of constructing a general model for multicellular living systems, a standard and efficient data representation capable of condensing the data while retaining essential features is required. The graph-based representation proposed in this study provides a possible solution. We note that while the nodal features and edges in the current analysis are purely geometrical, given the data we have at hand, important biological edges such as cell junctions and even long-ranged neurological connections can be further included in GNN models. Nevertheless, the positions of the cell nuclei can be used as the ``backbone" of multicellular data, to which multi-omics data can be attached. While ideally dynamic multi-omic experiments can be performed in one round with multi-channel live staining, practically multiple experiments performed on one biological sample typically require multiple runs under multiple microscopes. Coordinate alignment can be first achieved using the coordinates of the cell nuclei. With the graph-based data representation and graph-based deep neural network, the pipeline proposed in this work provides a possible solution for systematically organizing multicellular data into a multicellular data bank, and further condensed into graph-based deep neural networks for general predictions.

At a smaller length scale, deep neural nets (e.g. AlphaFold) have successfully uncovered the folded structure of ``graphs" of proteins~\cite{jumper2021highly}. In some recent studies, GNN models are shown to generate useful predictions within spatial transcriptomic datasets on fixed tissue samples~\cite{vinas2023hypergraph,hu2024unsupervised}. Nevertheless, it is still difficult to extend these data-driven efforts to dynamic time-lapsed video datasets of developmental processes consisting of a large number of cells evolving rapidly, with appropriate considerations of physical principles. We propose that it is possible to train a large model for general predictions of dynamic multicellular organization processes. The graph-based deep neural networks provide an excellent option that is capable of concatenating multi-omic biological inputs and outputs in an extremely flexible way. It will be interesting to further include genetics, proteomics, and other {\it in situ} multi-omic data as nodal features, and cell-cell junctions, mechanical interactions, and other interactions as edge features; these multi-omics graphs can then be provided to a graph neural network for prediction of missing or future features of these multicellular graphs evolving over time. Hence, we believe that the framework holds great potential for organizing fast-increasing multicellular data, and provides a possible solution to construct LMGM.

\clearpage
\subsection{Data acquisition and pre-processing procedures}

\subsubsection{Experimental dataset}
\begin{figure*}[h]
    \centering
    \includegraphics[width=0.45\textwidth]{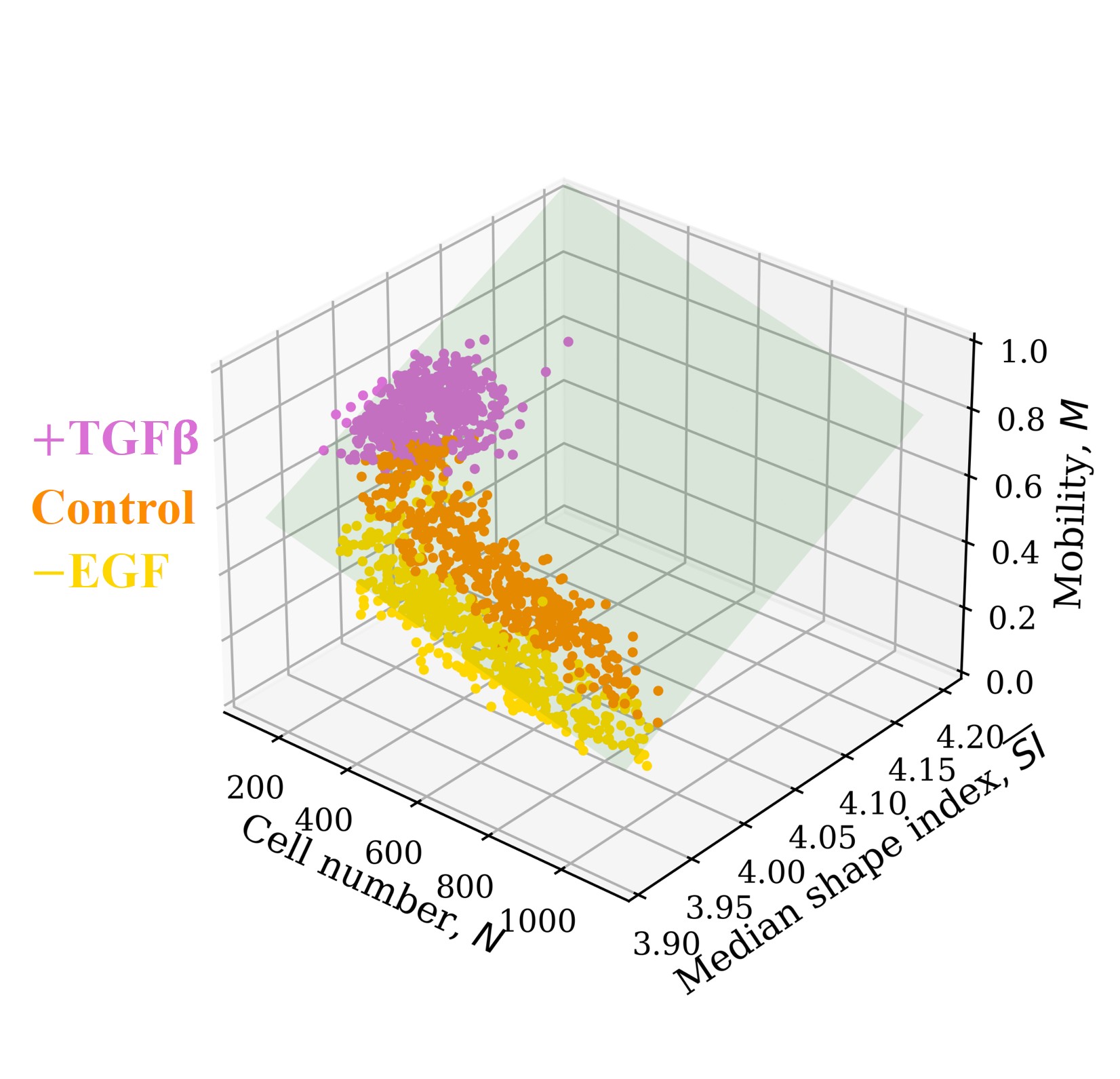}
    \caption{The mobility $M$ distribution of the full experimental dataset with $K=1440$ graphs. $M$ at lag time $\tau = 75$ minutes is shown here. Each dot corresponds to a graph. Note that all the graphs are treated equally and mixed in the datasets, and any information about the experimental condition, including the treatment type information (+TGF$\beta$, Control, -EGF), is not provided to the model. The green plane shows a fitted 2-variable linear regression model.}
    \label{fig:ExpRawDistribution}
\end{figure*}

\textit{Cell culture and imaging procedures.}---
Non-tumorigenic human breast epithelial cell line MCF-10A is used to create the experimental dataset. Cells are cultured at \SI{37}{\degreeCelsius} and 5\% $\mathrm{CO_2}$, with the full media of Dulbecco's modified Eagle's medium/F-12 supplemented with 5\% horse serum, \SI{20}{ng\,ml^{-1}} epidermal growth factor (EGF), \SI{0.5}{\micro\gram\, ml^{-1}} hydrocortisone, \SI{100}{ng\,ml^{-1}} cholera toxin, \SI{10}{\micro\gram\,ml^{-1}} insulin, and 1\% penicillin and streptomycin. 7 days before imaging, the parental cells are divided into two batches, with one batch still cultured in the full growth media, and the other batch cultured with the addition of \SI{5}{ng\,ml^{-1}} TGF$\beta$1 (PeproTech, No. 100-21). 1 day before imaging, cells are seeded on collagen-coated glass-bottom 96-well plates in three groups. Group A is the parental cells cultured in full growth media, and it will be used as the control group during imaging. Group B is also the parental cells cultured in full growth media, and it will be starved of EGF during imaging (to be treated with growth media without EGF). Group C is the batch that has been treated with TGF$\beta$1, and it is maintained with the same TGF$\beta$1 treatment throughout the experiment. Each group is seeded at 4 cell number densities at 15,000, 20,000, 25,000, and 30,000 cells per well. Around 2 hours before imaging, the cells are washed 2 times with serum-free media. Then the full growth media is added to group A, growth media without EGF is added to group B, and full growth media supplemented with \SI{5}{ng\,ml^{-1}} TGF$\beta$1 is added to group C, with all supplemented with SPY-650-DNA for cell tracking. 20 distinct locations of each condition are imaged simultaneously at 3 minutes per frame for 16 hours, with a field of view of \SI{802}{\micro\meter}$\times$\SI{599}{\micro\meter} at $1382 \times 1032$ pixels. In total, this results in 240 videos (each with 320 frames) covering a wide range of cell mobility (Main text, Fig.~2). Consistent with previous experimental studies~\cite{leggett2019motility,seton2004cooperation,hosseini2020emt}, the EGF withdrawal groups in general show reduced cell motility, and TGF$\beta$ treated groups show increased cell motility, lower proliferation and more irregular shapes as indicated by large cell shape index (Fig.~\ref{fig:ExpRawDistribution}).

\textit{Cell tracking.}---
Before tracking, the video resolutions are downsampled by 2, resulting in $691 \times 516$ pixels (resolution is 1.16 \SI{}{\micro\meter} per pixel). Automatic cell tracking is performed with a customized MATLAB script adapted from the particle tracking algorithm~\cite{crocker1996methods}. The tracking of each input snapshot is carefully examined and corrected using TrackMate~\cite{ershov2022trackmate}.

\textit{Frame selection.}---
We select the middle section of the videos for optimal video stability and tracking quality. Frames 100, 120, 140, 160, 180 and 200 from each video are selected to construct the whole dataset, each as one data entry. The frames are sparsely selected to reduce the correlations among the graphs. This results in a total of $K=1440$ graphs as inputs.

\textit{Normalization.}---
The characteristic area is calculated as $a_c = \frac{A_{FOV}}{\langle N \rangle}$, with the total area of the field of view $A_{FOV}=691\times 516 \, \mathrm{pixels}^2$ and $\langle N \rangle=\frac{1}{K} \sum_{k}^{K} N_k$ the average cell number per field of view, where $K = 1440$ is the total number of graphs in the full dataset. The length scale $a_c^{1/2} = 21.2 \text{ pixel}$ (\SI{24.6}{\micro\meter}) in this case.

The normalized cell coordinates are calculated as 
\begin{equation}
\mathbf{r}_i^* = a_c^{-1/2}\mathbf{r}_i\,.
\end{equation}
The field of view after normalization has a dimension of $x \in [0,32.5]$ and $y \in [0,24.3]$.

\textit{Calculating mobility.}---
Around each selected frame $t_s$, the mobility is defined as
\begin{equation}
    M = \langle \triangle {r^*}^2(\tau) \rangle ^{\frac{1}{2}},
\end{equation}
where the dimensionless mean squared displacement $\langle \triangle {r^*}^2(\tau) \rangle$ is calculated as
\begin{equation}
    \langle \triangle {r^*}^2(\tau) \rangle 
    = \mathrm{Var} \left( \mathbf{r}^*_i(t+\tau) - \mathbf{r}^*_i(t) \right) \big|_{i,t} \, ,
\end{equation}
with $\mathrm{Var()\big|_{i,t}}$ the variance taken both among cells and within the frame range $t \in [t_s-w, t_s+w]$, with $ w = 60$ frames (3 hours).

\textit{Input graphs.}---
For each frame, Delaunay tessellation is first performed to generate a list of edges. Then Voronoi tessellation is performed. To clean the distorted cells and edges at the boundary of the field of view, we select a region of interest (ROI) $x \in [2,30]$ and $y \in [2,22]$, and only keep nodes that have all their associated vertices within the ROI.

\textit{Train and validation datasets.}---
The full dataset contains a total of 12 conditions from 3 treatment groups and 4 seeding densities each at 20 independent locations, among which we use 18 locations for the train set, and 2 locations for the validation set. The mobility distribution of the 3 treatment groups as a function of cell number $N$ in the ROI and median shape index $\overline{SI}$ of the full dataset consisting of $K=1440$ graphs is shown in Fig.~\ref{fig:ExpRawDistribution}. All the graphs are treated equally and mixed in the datasets, and any information about the experimental condition is not provided to the GNN models.

\textit{Baseline model.}---
Over the full dataset with $K=1440$ graphs, the correlation between cell number $N$ in the ROI and the mobility $M$ is $-0.5728$, and the correlation between the median cell shape index $\overline{SI}$ and the mobility $M$ is $0.5599$. The baseline performance shown in Fig.~4(a) in the main text is determined using a two-variable linear regression model of the mobility $M$ against cell number $N$ in the ROI and the median cell shape index $\overline{SI}$ over the whole dataset, and the fitted plane is shown in Fig.~\ref{fig:ExpRawDistribution}, with a correlation of $\rho = 0.6173$.

Note that even without the edge information, a neural network consisting of multiple convolutional layers should still be able to consider higher-order statistic moments and nonlinear relations. Indeed, without the edge information, the $(a, p)$ group still reaches a correlation of $0.8134 \pm 0.0399$, significantly outperforming the two-variable linear regression model [main text, Fig.~4(a)].

\clearpage
\subsubsection{Synthetic dataset}

\textit{Basic information.}---
The Self-propelled Voronoi (SPV) simulations are performed following the procedure as described in~\cite{bi2016motility}. Briefly, the free energy is given by
\begin{equation}
    E = \Sigma_{i=1}^N \left[  K_A (A_i-A_0)^2 + K_P (P_i-P_0)^2 \right],
\end{equation}
where $A_i$ and $P_i$ are the area and perimeter of the Voronoi cell $i (i=1,2,..., N)$, the coefficients $A_0$ and $P_0$ are the target cell area and perimeter, and the coefficients $K_A$ and $K_P$ are the area and perimeter moduli.

The equation of motion is given by
\begin{equation}
    \frac{d \mathbf{r}_i}{dt} = \mu \mathbf{F}_i + v_0 \mathbf{n}_i,
\end{equation}
where $\mathbf{r}_i$ is the Cartesian coordinate of the cell $i$, $\mathbf{n}_i = (\cos{\theta},\sin{\theta})$ is the cell polarity vector, $\mathbf{F}_i = -\nabla_i E$ is the force, the coefficient $v_0$ is the self-propulsion speed, and the coefficient $\mu$ is a drag coefficient.

The cell polarity follows
\begin{equation}
    \partial_t \theta_i = \eta_i(t), \langle \eta_i(t) \eta_i(t') \rangle = 2 D_r \delta(t-t') \delta_{ij},
\end{equation}
where $\eta_i(t)$ is the white-noise process with zero mean and variance $2D_r$.

\textit{Nondimensionalization.}---
Same as in~\cite{bi2016motility}, we consider a confluent cell monolayer where the average cell area $\langle A_i \rangle = A_0$. Choosing $\sqrt{A_0}$ as the length scale and $\frac{1}{\mu K_A A_0}$ as the time scale and nondimensionalize the equations, the system has 3 independent parameters, which are the target cell shape index $p_0$ ($p_0 = \frac{P_0}{\sqrt{A_0}}$), the self-propulsion speed $v_0$ and the angular diffusion coefficient $D_r$.

\textit{The dataset.}---
In this study, we use a dataset at $D_r = 2$ containing simulations at 462 distinct state points covering a broad range of cell target shape index $p_0$ and self-propulsion speed $v_0$. The simulations are performed at a time step size of $\Delta t = 0.1$ and the results are recorded every 5 time steps (frame rate is 0.5 dimensionless time per frame). Each simulation contains 4,000 frames (total dimensionless time is 2,000) of $N = 400$ cells within a simulation box of dimension $x \in [0,20]$ and $y \in [0,20]$, and therefore the characteristic length scale is $a_c^{1/2} = 1$ in the simulation.

\textit{Frame selection.}---
Unless otherwise noted, from the total 462 state points we randomly select 30 state points for training as indicated in Fig.~3(d) in the main text, and we select frame 500 to frame 3500 with a step size of 500, resulting in 7 frames per state point. The frames are sparsely selected to reduce the correlation among the graphs. Model performance of selecting a different number of frames per state points or selecting different state points for training are shown in Table.~\ref{tab:NumFrame}, Table.~\ref{tab:NumState} and Fig.~\ref{fig:SimuChangeNumState}. 

\textit{Calculating mobility.}---
The cell mobility $M$ is calculated the same as in the experimental dataset with $w=100$.

\textit{Input graphs.}---
The input graphs are constructed in the same way as in the experimental dataset. We select ROI as $x \in [0,20]$ and $y \in [0,20]$.

\textit{Output normalization for training.}---
The synthetic dataset covers a much broader range of cell mobility, with the maximal $M$ on the order of 10 in a highly fluid-like system. For the benefit of using consistent hyperparameters in training the neural network, it is helpful to rescale the output around 1. Therefore, we rescale $M$ as $\hat{M} = M/M_c$, with $M_c=10$ when providing the data to the GNN model. That is, the direct output of the GNN is $\hat{M}$, and we scale $\hat{M}$ back to its original range by $M_{NN} = M_c \hat{M}$ for all post-processing and visualization.

\begin{figure*}[h]
    \centering
    \includegraphics[width=0.35\textwidth]{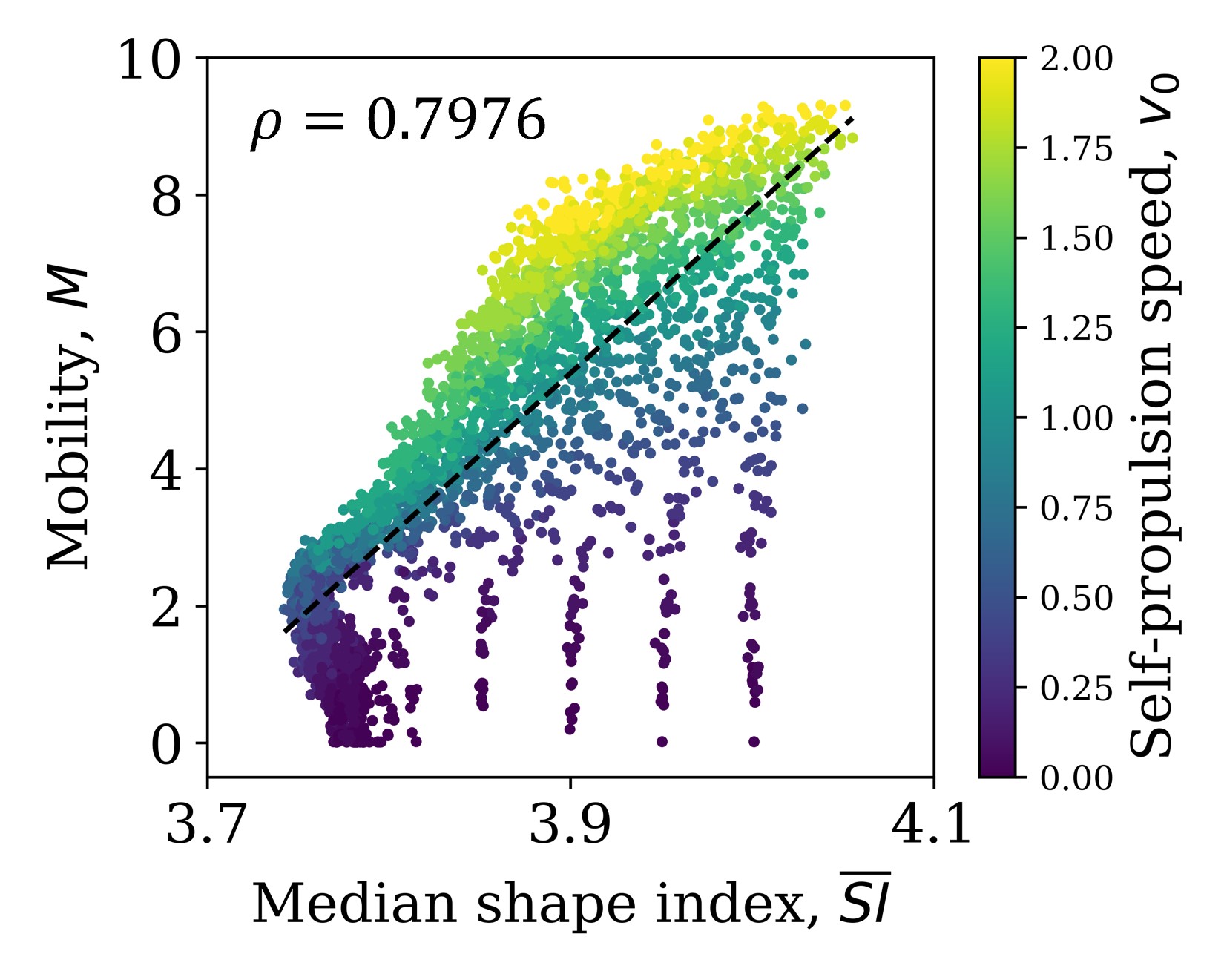}
    \caption{The mobility distribution of the synthetic dataset. $M$ at the dimensionless lag time $\tau=10$ is shown here. The dashed line represents a fitted linear regression model.}
    \label{fig:SimuRawDistribution}
\end{figure*}

\textit{Baseline model.}---
The baseline shown in Fig.~4(b) in the main text is determined using a linear regression model of the mobility $M$ against the median cell shape index $\overline{SI}$ over the whole dataset. The fitted line is shown in Fig.~\ref{fig:SimuRawDistribution} with a correlation of $\rho = 0.7976$.

\clearpage
\subsection{Additional results}
\subsubsection{Performance at different time scales}

\begin{figure*}[h]
    \centering
    \includegraphics[width=0.85\textwidth]{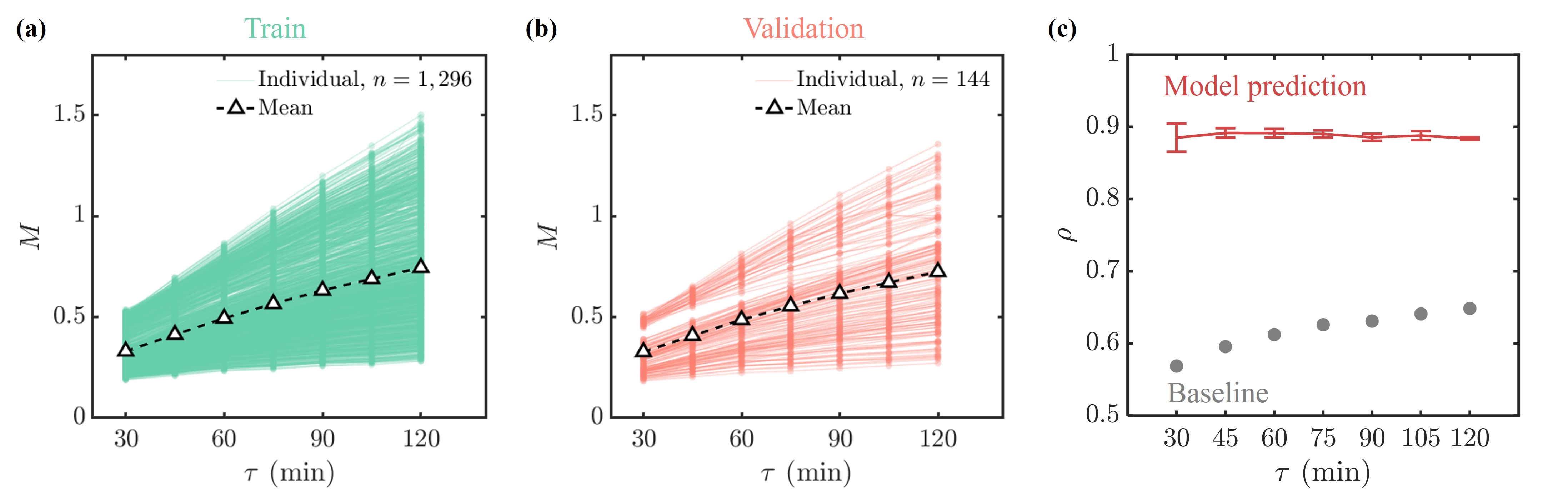}
    \caption{For the experimental dataset, (a) mobility $M(\tau)$ at different time scales in the training set, and (b) mobility $M(\tau)$ at different time scales in the validation set. (c) The model performance at different time scales. The train and validation set are the same as in Fig.2. The models are initialized with 3 different random seeds and the error bar shows the standard deviation. The baseline is determined by the Pearson correlation between $M$ and the median of cell shape index $\overline{SI}$ at time scale $\tau$.}
    \label{fig:ExpLagTime}
\end{figure*}

In both experimental and synthetic datasets, the selection of time scales has minor effects on the performance of the GNN model (Fig.~\ref{fig:ExpLagTime} and Fig.~\ref{fig:SimuLagTime}). This perhaps suggests that from the graphs the models extract structural features reflecting differences from condition to condition, instead of identifying instantaneous cell-cell rearrangements; for example, in the synthetic dataset, these could be some representation similar to $(p_0, v_0)$ which then could be used to interpolate the mobility quite easily since the landscape in the $(p_0, v_0, M)$ space is relatively smooth, although it is also possible that the models trained on different time scales have captured different structural features while achieving similar performance. In jamming transitions in the experimental cell monolayers, the ``internal" state analogous to $(p_0, v_0)$ remains an elusive concept, however, the output of the last graph convolutional layer before the final linear layer of a trained GNN model could perhaps be regarded as a low-dimensional structural representation of the cell monolayers, relevant to their jamming dynamics. For future studies, it will be interesting to visualize these hidden features and analyze their changes under biophysical and biochemical perturbations.

\clearpage
\begin{figure*}[h]
    \centering
    \includegraphics[width=0.85\textwidth]{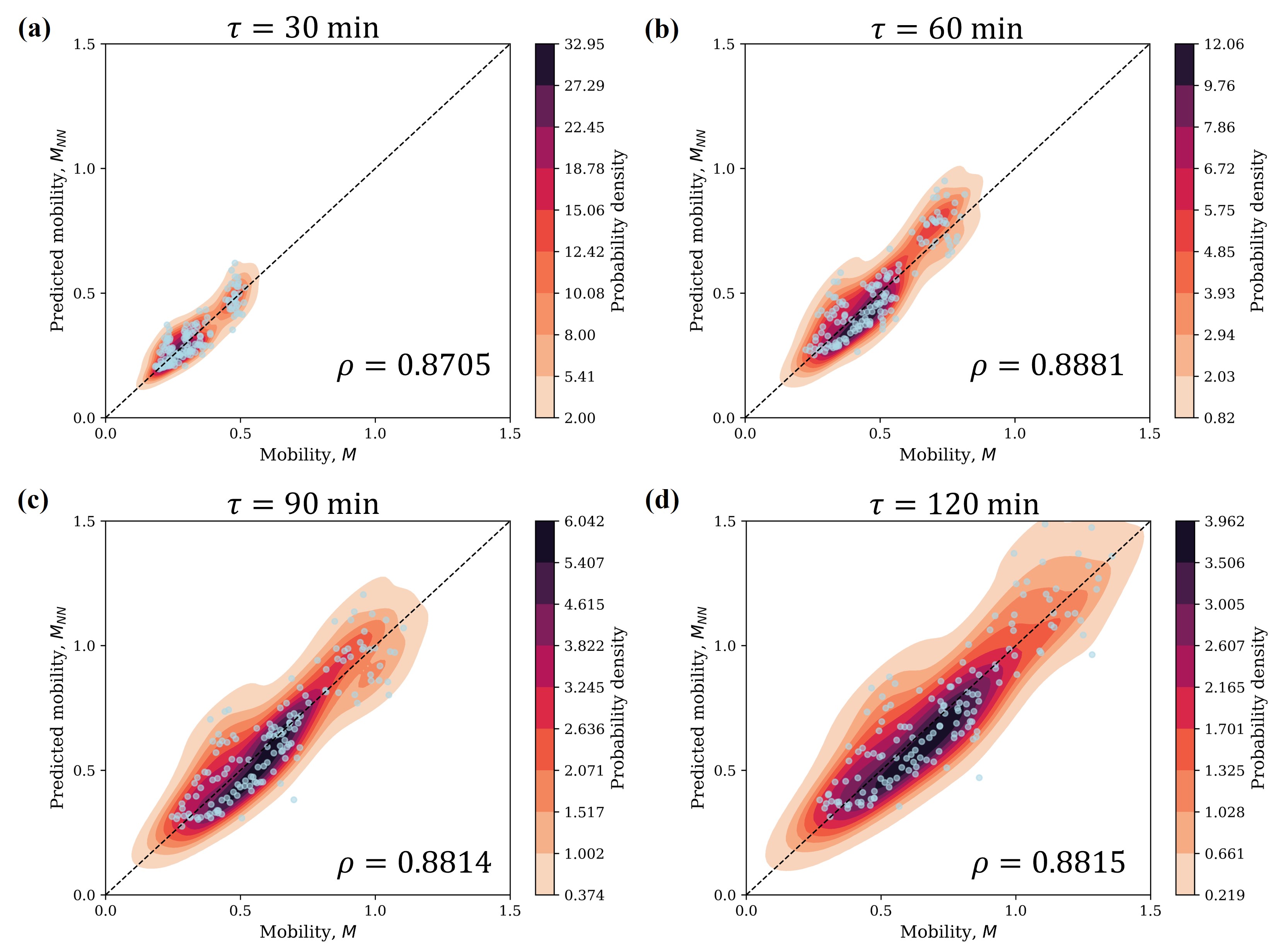}
    \caption{GNN predictions vs. ground truth at different time scales $\tau$ in the experimental dataset. (a) $\tau=30$ min. (b) $\tau=60$ min. (c) $\tau=90$ min. (d) $\tau=120$ min.}
    \label{fig:ExpLagTimeDemo}
\end{figure*}

\clearpage
\begin{figure*}[h]
    \centering
    \includegraphics[width=0.85\textwidth]{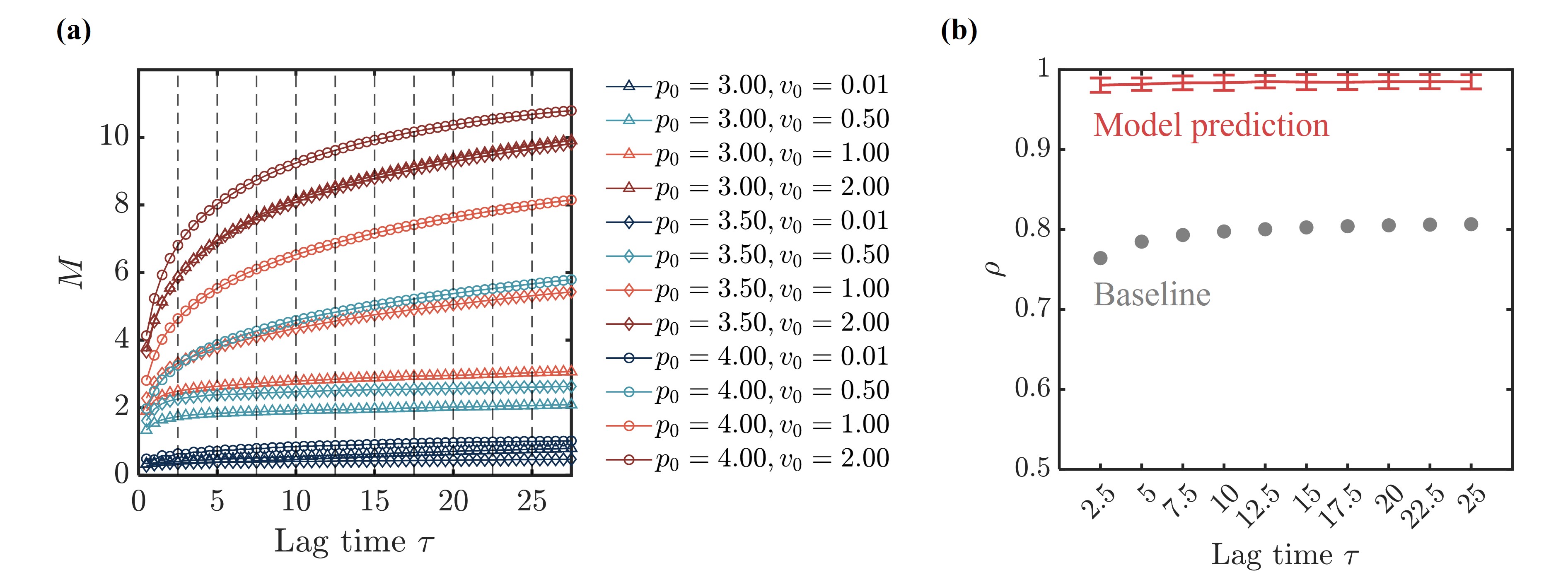}
    \caption{(a) Examples of mobility $M(\tau)$ at different time scales in the synthetic dataset. The dashed line marks the time scales on which the models are trained and validated. (b) The model performance. 3 different train sets of 30 state points are randomly selected, on which separate models are trained. The error bar shows the standard deviation. The baseline is determined by the Pearson correlation between $M$ and the median of cell shape index $\overline{SI}$ at the dimensionless time scale $\tau$.}
    \label{fig:SimuLagTime}
\end{figure*}

\clearpage
\begin{figure*}[h]
    \centering
    \includegraphics[width=0.98\textwidth]{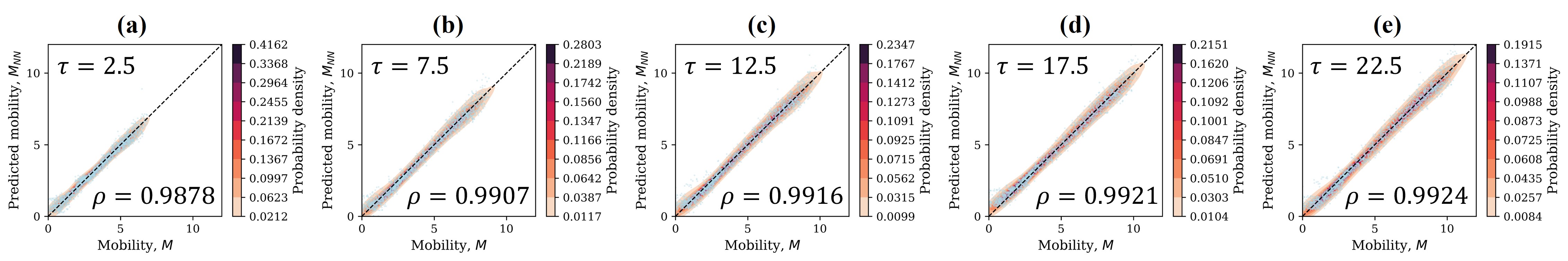}
    \caption{GNN predictions vs. ground truth at different dimensionless time scales $\tau$ in the synthetic dataset. (a) $\tau=2.5$. (b) $\tau=7.5$. (c) $\tau=12.5$. (d) $\tau=17.5$. (e) $\tau=22.5$.}
    \label{fig:SimuLagTimeDemo}
\end{figure*}

\clearpage
\subsubsection{Effect of model hyperparameters}

\begin{table}[h]
    \centering
    \begin{tabular}{ccccc}
    \hline
    Layer & Channel & Batch size & Learning rate & $\rho$  \\ \hline 
    5 & 36 & 128 & 1e-05 &       0.8902$\pm$0.0050       \\ 
    2 & 36 & 128 & 1e-05 &       0.8660$\pm$0.0114       \\
    3 & 36 & 128 & 1e-05 &       0.8913$\pm$0.0042       \\ 
    4 & 36 & 128 & 1e-05 &       0.8857$\pm$0.0046       \\ 
    6 & 36 & 128 & 1e-05 &       0.8850$\pm$0.0067       \\ 
    7 & 36 & 128 & 1e-05 &       0.8767$\pm$0.0079       \\
    
    5 & 12 & 128 & 1e-05 &       0.8494$\pm$0.0236       \\ 
    5 & 24 & 128 & 1e-05 &       0.8893$\pm$0.0047       \\ 
    5 & 48 & 128 & 1e-05 &       0.8818$\pm$0.0007       \\ 
    5 & 60 & 128 & 1e-05 &       0.8837$\pm$0.0091       \\ 
    
    5 & 36 & 32 & 1e-05 &        0.8700$\pm$0.0128       \\ 
    5 & 36 & 64 & 1e-05 &        0.8812$\pm$0.0026       \\ 
    5 & 36 & 256 & 1e-05 &       0.8863$\pm$0.0080       \\
    
    5 & 36 & 128 & 1e-04 &       0.8018$\pm$0.0341       \\ 
    5 & 36 & 128 & 1e-06 &       0.8087$\pm$0.0814       \\
    \hline
    \end{tabular}
    \caption{Effect of hyperparameters. The same train and validation sets are used as Fig.~2 in the main text. Each group is repeated 3 times with different random-seed initialization.}
    \label{tab:HyperParam}
\end{table}

\subsubsection{Effect of dataset size}

\begin{table}[h]
    \centering
    \begin{tabular}{c|c}
    \hline
    State points         & $\rho$       \\ \hline 
    5                    &  0.9159 $\pm$ 0.0445      \\
    10                   &  0.9552 $\pm$ 0.0264      \\
    15                   &  0.9765 $\pm$ 0.0041      \\
    20                   &  0.9760 $\pm$ 0.0146      \\
    25                   &  0.9798 $\pm$ 0.0128      \\
    30                   &  0.9837 $\pm$ 0.0096      \\
    35                   &  0.9840 $\pm$ 0.0097      \\
    40                   &  0.9900 $\pm$ 0.0034      \\
    45                   &  0.9902 $\pm$ 0.0025      \\
    50                   &  0.9909 $\pm$ 0.0022      \\
    \hline
    \end{tabular}
    \caption{Performance on the synthetic dataset with different number of state points randomly selected for training. Each condition is repeated 3 times with a different random selection of state points for training.}
    \label{tab:NumStateRand}
\end{table}

\begin{table}[h]
    \centering
    \begin{tabular}{c|c}
    \hline
    Number of frames per state points       & $\rho$       \\ \hline 
    2                                       & 0.9777 $\pm$ 0.0050    \\
    7                                       & 0.9837 $\pm$ 0.0096    \\
    \hline
    \end{tabular}
    \caption{Performance on the synthetic dataset with different number of frames per state point provided in the training set. The raw dataset contains 4000 frames of steady-state configuration; here for the 2-frame group, the frames with frame index $\in \{500,3500\}$ are provided each as one separate data entry in the full dataset, and for the 7-frame group, the frames with frame index $\in \{500, 1000, 1500, 2000, 2500, 3000, 3500\}$ are provided each as one separate data entry in the full dataset. For both, 30 randomly selected state points are provided for training. For each group, 3 individual models are trained, each on a different randomly selected train set, and the models are validated on the full datasets consisting of 462 state points.}
    \label{tab:NumFrame}
\end{table}

\begin{table}[h]
    \centering
    \begin{tabular}{c|c}
    \hline
    State points        & $\rho$       \\ \hline 
    $2\times 2$         &  0.9267 $\pm$ 0.0357      \\
    $3\times 3$         &  0.9835 $\pm$ 0.0045      \\
    $4\times 4$         &  0.9877 $\pm$ 0.0030      \\
    $5\times 5$         &  0.9909 $\pm$ 0.0005      \\
    $6\times 6$         &  0.9932 $\pm$ 0.0010      \\
    \hline
    \end{tabular}
    \caption{Performance on the synthetic dataset with different state points on regular mesh grids provided in the training set. The corresponding state points $(p_0,v_0)$ provided in the training set are marked in Fig.~\ref{fig:SimuChangeNumState}. Each condition is repeated 3 times with different random seeds.}
    \label{tab:NumState}
\end{table}

\begin{figure*}[h]
    \centering
    \includegraphics[width=0.6\textwidth]{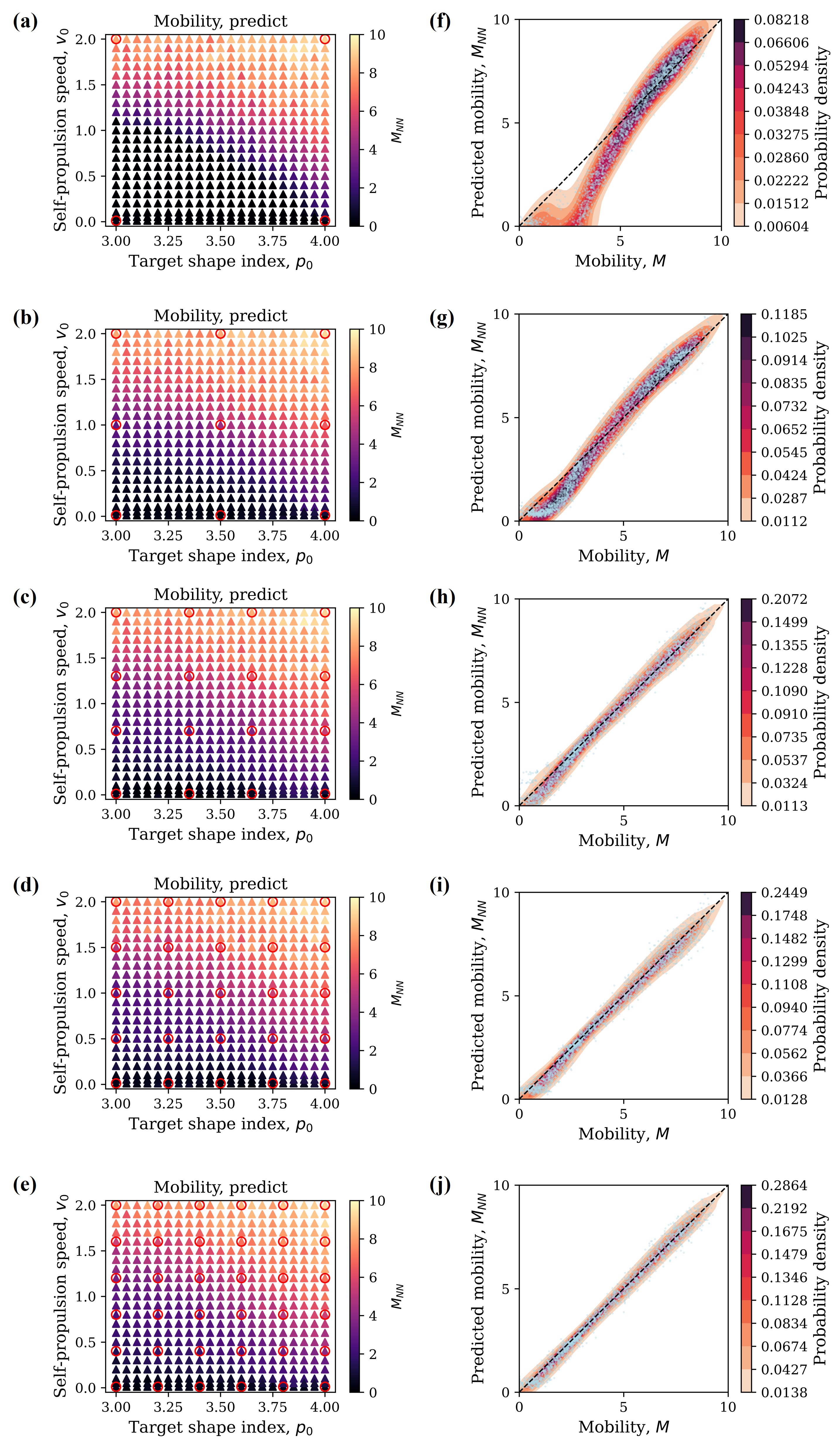}
    \caption{The effect of states provided for training. (a-e) The predicted mobility landscape. The red markers indicate the states provided for training. (f-j) Prediction vs. ground truth. The color map indicates probability density and each scatter point represents one input graph.}
    \label{fig:SimuChangeNumState}
\end{figure*}

\clearpage
\bibliography{bibliography}